\documentclass{article}

\usepackage{PRIMEarxiv}

\usepackage[utf8]{inputenc} 
\usepackage[T1]{fontenc}    
\usepackage{hyperref}       
\usepackage{url}            
\usepackage{booktabs}       
\usepackage{amsfonts}       
\usepackage{nicefrac}       
\usepackage{microtype}      
\usepackage{lipsum}
\usepackage{fancyhdr}       
\usepackage{graphicx}       
\graphicspath{{media/}}     

\title{AI Literature Review Suite}

\author{
  David A. Tovar \\
  Department of Psychology \\
  Vanderbilt University \\
  Nashville, TN\\
  \texttt{david.tovar@vanderbilt.edu} \\
}

\begin{document}
\maketitle

\begin{abstract}
The process of conducting literature reviews is often time-consuming and labor-intensive. To streamline this process, I present an AI Literature Review Suite that integrates several functionalities to provide a comprehensive literature review. This tool leverages the power of open access science, large language models (LLMs) and natural language processing to enable the searching, downloading, and organizing of PDF files, as well as extracting content from articles. Semantic search queries are used for data retrieval, while text embeddings and summarization using LLMs present succinct literature reviews. Interaction with PDFs is enhanced through a user-friendly graphical user interface (GUI). The suite also features integrated programs for bibliographic organization, interaction and query, and literature review summaries. This tool presents a robust solution to automate and optimize the process of literature review in academic and industrial research.
\end{abstract}


\section{Introduction}

In academic and industry research, literature reviews serve as the cornerstone of extensive comprehension and exploration of any given topic. Traditional manual processes of literature reviews are, however, characterized by time-consuming and labor-intensive tasks. These tasks often include sifting through volumes of academic papers, manually downloading and organizing relevant ones, reading and summarizing these papers, and finally, synthesizing the information into a cohesive narrative. The sheer magnitude of academic papers published daily and the complexity of most research topics compound this issue further. Consequently, there has been a growing need for more efficient tools that can automate and streamline the literature review process, thereby enabling researchers to focus more on knowledge synthesis and less on the logistical aspects of conducting literature reviews.

To address this need, I introduce the AI Literature Review Suite, a comprehensive suite of integrated programs for conducting literature reviews efficiently and accurately. This tool capitalizes on the advancements in machine learning and natural language processing to automate several tasks involved in the literature review process. It includes features such as searching, downloading, and organizing PDF files, extracting content from articles, performing semantic search queries, summarizing literature reviews, and providing a user-friendly interface for interacting with PDFs. By automating these tasks, the tool greatly reduces the amount of time and effort required to conduct comprehensive literature reviews.

The suite comprises several integrated programs, each designed to perform specific tasks. The \texttt{PDF Search} program, for instance, interacts with the CORE API to search for and download scholarly articles based on user-provided parameters. The \texttt{PDF Extraction} program serves as a bibliographic tool that leverages the CORE API \cite{core_api_website} and the CrossRef RESTful API \cite{crossref_website} to download and organize articles along with their references and citations based on DOIs. The \texttt{PDF Chat} program borrows from a previous solution \cite{github_website}, but adds number of features including a graphical user interface (GUI) for interaction, ability to ask specific and general questions, and saving conversation in a word document for future reference. The \texttt{Literature Review Table} program is an automatic literature review tool that processes multiple PDFs, performs semantic search queries, and generates comprehensive responses. The \texttt{Literature Synthesis} program uses semantic embeddings \cite{yang_multilingual_2019,cer_universal_2018} and large language models \cite{touvron_llama_2023,openai_gpt-4_2023} to automatically create detailed summaries from multiple text entries.

This paper aims to provide a detailed overview of the AI Literature Review Suite, including its design, functionalities, and potential applications in academic and industrial research. I also discuss the integrated programs that form the backbone of the tool, their features, and how they work together to streamline the literature review process. The primary objective is to illustrate how this tool can enhance efficiency and quality in conducting literature reviews, ultimately catalyzing knowledge discovery in various research fields.

\section{Results}
\label{sec:headings}
The architecture of the AI Literature Review Suite is underpinned by a principle of modularity, designed with the intention of offering researchers the flexibility to choose how much or how little of the tool they wish to use. The suite is structured into three main modules: "Knowledge Gathering," "Knowledge Extraction," and "Knowledge Synthesis." Each of these modules encapsulates a fundamental aspect of the literature review process, and together, they present a holistic approach to conducting literature reviews.

The "Knowledge Gathering" module provides functionalities that facilitate the sourcing and organization of relevant academic papers. Tools like \texttt{PDF Search} and \texttt{PDF Extraction} are incorporated in this module, allowing researchers to search, download, and neatly organize articles based on specified parameters. Researchers can leverage these tools to build a comprehensive repository of relevant literature effortlessly. If a researcher wishes only to use the suite for these tasks, they are entirely at liberty to do so.

The next module, "Knowledge Extraction," aids in the extraction and processing of content from the gathered articles. Here, the \texttt{PDF Chat} tool and  comes into play, offering functionalities like text extraction and semantic search for enhanced interaction with the academic papers. Researchers can ask document-specific questions and receive accurate answers, aiding in a thorough understanding of the papers.

The final module, Knowledge Synthesis, utilizes and to create concise summaries of the content extracted from the articles and synthesize a cohesive narrative that encapsulates the central theme of the literature review. This module essentially aids in transforming the raw, extracted information into a consumable format, easing the process of knowledge assimilation.

\begin{figure}
  \centering
  \includegraphics[width=0.8\textwidth]{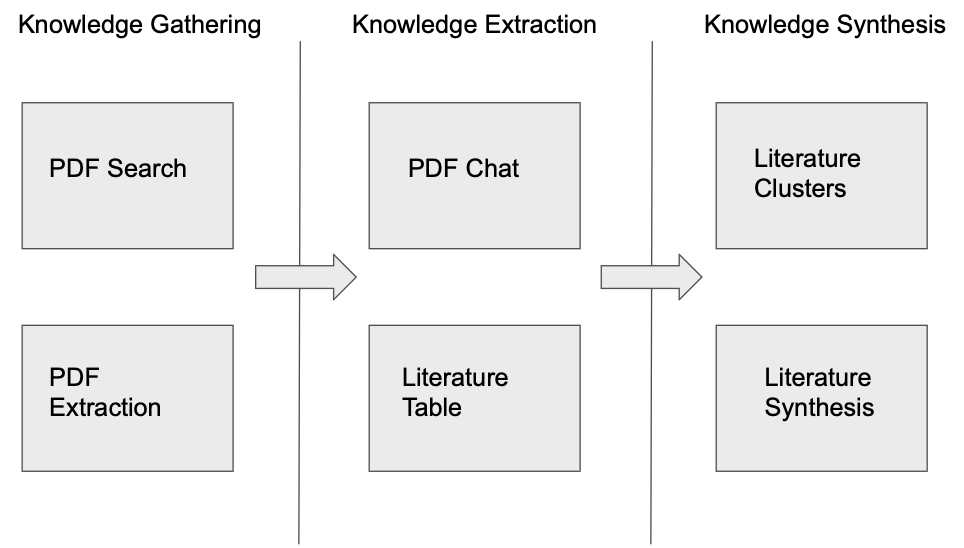}
  \caption{AI Literature Review Suite Schematic and Modules}
  \label{fig:fig1}
\end{figure}

Researchers have the flexibility to navigate these modules either interactively or in an automated fashion, choosing to manually guide the process for more control or let the suite's robust automation handle the process end-to-end. This versatility in usage, coupled with the suite's ability to run on consumer laptops and open source Python \cite{python3} software  positions the AI Literature Review Suite as a significant aid to researchers, assisting in conducting effective and efficient literature reviews.

\subsection{Graphic User Interface}

The Graphic User Interface (GUI) is designed to facilitate user interaction with the suite and offer a streamlined user experience. The interface is implemented using a modern, cross-platform framework in Python that supporting Windows, macOS, and Linux. The GUI is divided into different sections corresponding to each integrated program or module, each equipped with a distinct set of controls and visualizations to guide researchers through the process with their own GUI which will be described in the sections below.

\begin{figure}
  \centering
  \includegraphics[width=0.8\textwidth]{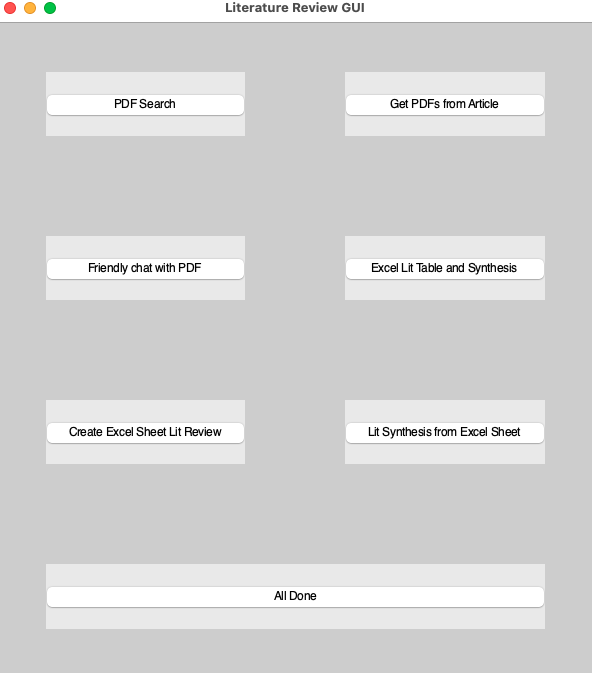}
  \caption{Graphic user interface with selections for each module }
  \label{fig:fig2}
\end{figure}

\subsection{PDF Search}
The PDF search module, a key component of the AI Literature Review Suite, leverages the capabilities of the CORE API \cite{core_api_website} to access a wide array of open-access articles, including those hosted on individual lab websites. This ensures an expansive literature search, enhancing the chances of retrieving all pertinent literature on a given topic. The module allows researchers to designate specific search parameters, such as topics, titles, authors, and publication years, promoting a tailored literature retrieval process.

\begin{figure}
  \centering
  \includegraphics[width=0.8\textwidth]{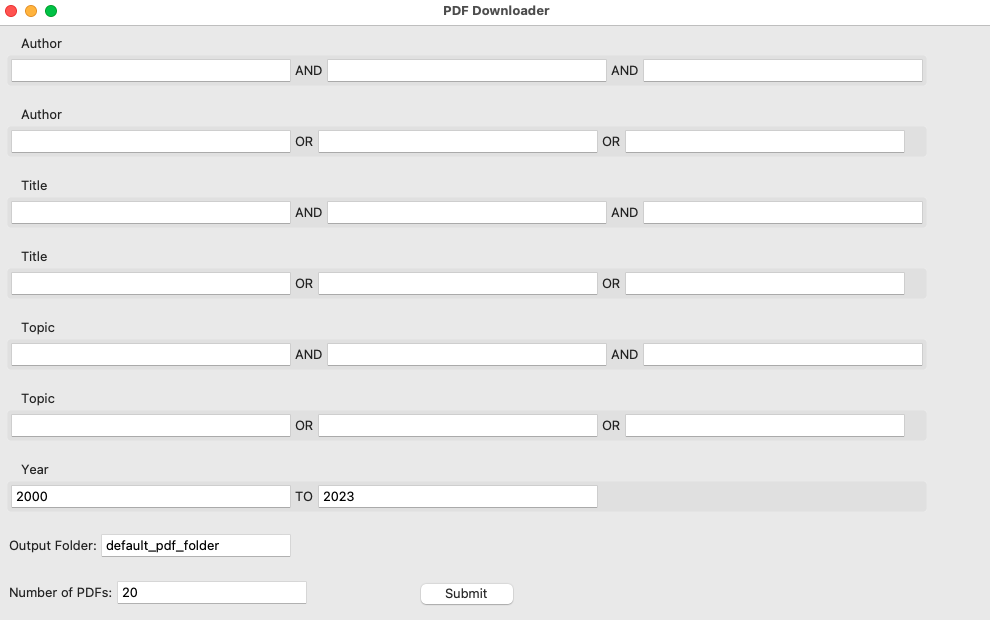}
  \caption{GUI for PDF search using CORE Database}
  \label{fig:fig3}
\end{figure}

Upon article retrieval, they are systematically stored in a user-specified folder with the citation in APA format, facilitating subsequent referencing. Concurrently, URL links are preserved in separate text files, pre-formatted for insertion as hyperlinks, optimizing accessibility. The module also documents articles not found in the CORE database in a separate text file with their authors, titles, and abstracts. This mechanism ensures that all potential information sources are accounted for in the search process.

\subsection{PDF Extraction}

The PDF Extraction offers a focused approach to literature extraction from selected PDFs. This module utilizes the CrossRef API \cite{crossref_website} and literature scanner python package \cite{donoghue2018lisc} to extract metadata, references, and citations from a PDF, thereby identifying valuable resources for a comprehensive literature review. The module searches the CORE API to acquire open-access PDFs that align with the extracted metadata. The PDFs are then saved in a user-specified directory, with citations presented in APA format, to streamline future referencing tasks. The module has an additional feature of categorizing and segregating citations and references into a subfolder. This classification enhances the accessibility and readability of the extracted data, which simplifies subsequent literature analysis tasks.

\subsection{PDF Chat}

The PDF Chat module is an interactive tool, designed to facilitate querying any selected PDF using a Large Language Model, such as GPT \cite{openai_gpt-4_2023} or LLaMA \cite{touvron_llama_2023}. The module allows researchers to inquire about the main message or request  the main results be presented in a numbered list format. This targeted questioning permits a precise extraction of the crux of a study, significantly augmenting the understanding of complex academic texts. A standout feature of this module is its flexibility. Researchers can choose to question as many or as few PDFs as needed, tailored to their individual research requirements. Importantly, the questions are limited to the context of the selected PDF captured through the semantic embedding models \cite{github_website, cer_universal_2018}, ensuring relevant and precise responses, mitigating the issue of hallucinations, a common concern with AI language models \cite{openai_gpt-4_2023, touvron_llama_2023}. However, the module still allows for the exploration of information outside the PDF. If researchers desire additional context or a broader understanding, they can ask general questions in the same dialogue. Lastly, the PDF Chat module records the entire conversation and stores it as a Word document. This feature allows researchers to refer back to the extracted information and the line of questioning, providing a valuable reference for further research analysis.

\begin{figure}
  \centering
  \includegraphics[width=0.8\textwidth]{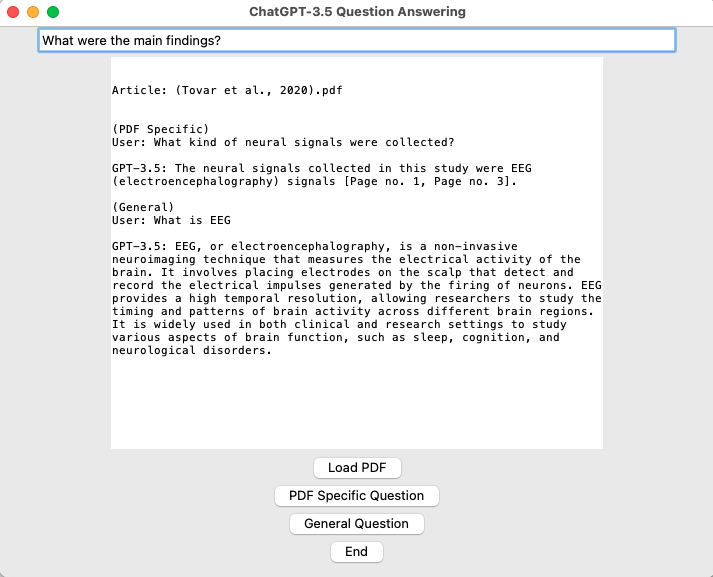}
  \caption{Example chat with PDF with specific and general questions}
  \label{fig:fig4}
\end{figure}

\subsection{Literature Table}

The Literature Table module represents a practical solution to efficiently manage and summarize a large volume of academic articles. It facilitates the creation of an organized Excel table from a folder of selected PDFs, where each row corresponds to an individual article. The table consists of columns representing key elements of an academic article. These include the APA in-text citation, providing a ready-to-use reference for future scholarly work, and the summaries of the Introduction, Methods, and Results sections of each article. The module also offers customization by allowing researchers to pose their own questions that replace the default queries for the introduction, methods, or results summaries. This feature enables targeted literature analysis, enabling researchers to quickly access the specific information they need.

These summaries are generated by a synergistic use of a semantic embedding model \cite{cer_universal_2018} and a Large Language Model \cite{openai_gpt-4_2023, touvron_llama_2023}, ensuring coherent, meaningful, and concise representations of the original text. Furthermore, in the case of subfolders, the Literature Table module creates separate Excel sheets for each subfolder within the main Excel file, ensuring a neatly organized output that mirrors the original file structure.

\begin{figure}
  \centering
  \includegraphics[width=1\textwidth]{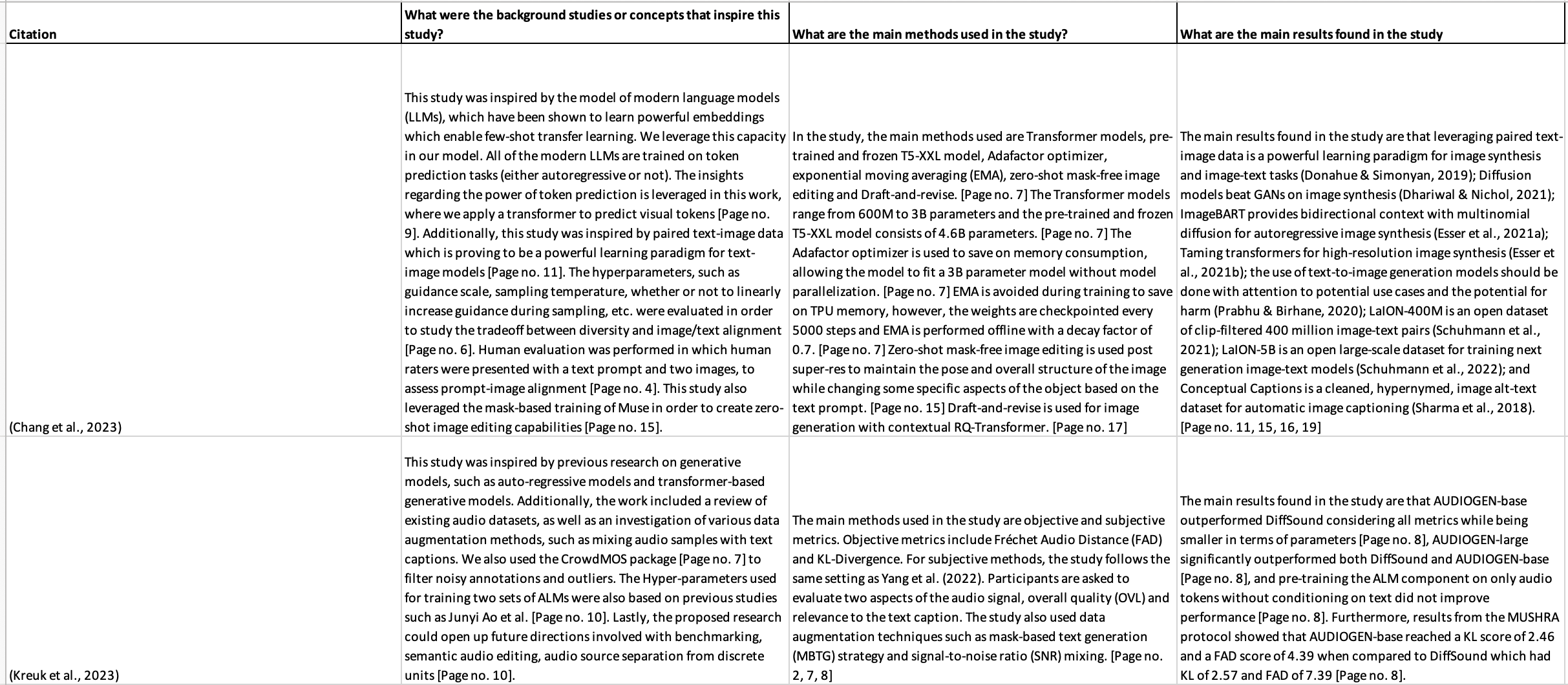}
  \caption{Example rows from literature table}
  \label{fig:fig5}
\end{figure}

\subsection{Literature Clusters}

The Literature Clusters module serves to intelligently categorize and group academic articles based on their content. This module processes each row of the previously created Literature Table, feeding them into a semantic embedding model \cite{cer_universal_2018}. It then uses K-nearest neighbors (K =5) to group the articles into clusters \cite{scikit-learn,harris2020array,reback2020pandas}, each typically containing five closely related works. This approach ensures that the clusters represent meaningful groupings within the literature, facilitating a deeper understanding of the nuanced areas of study within a broader research field.

Upon completion of the clustering process, the module generates a Word document that groups the Excel rows according to their respective clusters. This document maintains the original Excel text, ensuring the preservation of the rich information extracted in the Literature Table module. This methodology provides researchers with a clear and organized representation of the literature landscape, helping them identify patterns, trends, and areas for further exploration.

\subsection{Literature Synthesis}

The Literature Synthesis module uses a  large language model to distill, compare, and contrast the grouped works obtained from the Literature Clusters module, creating synthesized paragraphs that capture key themes, similarities, and differences. Each section and sentence within the summaries have the appropriate APA citations that the user can go refer back to. For this part of the AI suite, it uses the advanced capabilities of GPT-4 \cite{openai_gpt-4_2023} to generate coherent, meaningful, and detailed syntheses of clustered literature. However, one of the key features of this module is its flexibility and adaptability. The system is designed to be modular and can easily be integrated with future open-source models, ensuring that it remains at the forefront of AI advancements and continues to provide high-quality literature synthesis.

\section{Discussion}

At a time when information is being generated at an unprecedented pace, the AI Literature Review Suite can have considerable impact. It particularly shines in rapidly evolving domains such as medicine, science, and engineering, where keeping up with the latest research findings is crucial. For medical practitioners, scientists, and engineers, this tool can substantially expedite the process of assimilating the latest research. It reduces the time and effort required for conducting comprehensive literature reviews, ensuring that vital findings are not overlooked. In medicine, it can directly impact patient care by enabling clinicians and policy makers to make more informed decisions. In the realms of science and engineering, it accelerates the iterative cycle of hypothesis generation, testing, and refinement, thereby spurring innovation and progress.

The suite is also designed to adapt and evolve with the fast-paced growth of artificial intelligence. It can integrate with different large language models, making it an adaptable tool in the dynamic landscape of AI research and application. This ensures its continued relevance and utility across various fields. However, while this tool offers many benefits, it is essential to view it as a first step in acquainting oneself with a topic rather than the definitive source of information. While the risk of losing some information in the text embedding process is mitigated through methods like different initialization seeds, critical matters always warrant a careful inspection of the literature. In summary, the AI Literature Review Suite stands as a potent ally for researchers, enhancing efficiency and quality of scholarly endeavors while promoting accelerated innovation and progress.

\section*{Github}
Github with the latest release : \href{https://github.com/datovar4/AI_Literature_Review_Suite}{AI Literature Review Suite} 

\section*{Acknowledgments}
Thank you to Ian Erkelens and Mark Wallace for feedback on early versions of the AI Literature Suite.

\bibliographystyle{unsrt}  
\bibliography{references}

\end{document}